# DRIVER ACCELERATOR DESIGN FOR THE 10 KW UPGRADE OF THE JEFFERSON LAB IR FEL

D. Douglas, S. V. Benson, G. A. Krafft, R. Li, L. Merminga, and B. C. Yunn
Thomas Jefferson National Accelerator Facility, Newport News, VA 23606, USA

*Abstract*

An upgrade of the Jefferson Lab IR FEL [1] is now under construction. It will provide 10 kW output light power in a wavelength range of 2–10 µm. The FEL will be driven by a modest sized 80–210 MeV, 10 mA energy-recovering CW superconducting RF (SRF) linac. Stringent phase space requirements at the wiggler, low beam energy, and high beam current subject the design to numerous constraints. These are imposed by the need for both transverse and longitudinal phase space management, the potential impact of collective phenomena (space charge, wakefields, beam break-up (BBU), and coherent synchrotron radiation (CSR)), and interactions between the FEL and the accelerator RF system. This report addresses these issues and presents an accelerator design solution meeting the requirements imposed by physical phenomena and operational necessities.

## 1 DESIGN REQUIREMENTS

The Jefferson Lab IR Upgrade FEL will be an evolutionary derivative of the JLab IR Demo FEL. As such, it retains the paradigm invoked for the earlier machine – that of a low peak, high average power wiggler-driven optical cavity resonator with an energy recovering SRF linear accelerator driver operating at high repetition rate. The 10 kW design goal will be achieved via an increase in both drive beam power (doubled current and quadrupled energy) and FEL extraction efficiency (from ½% to 1%), and will be accommodated through use of an $R^5$ optical cavity geometry [2].

This system concept imposes two fundamental requirements on the driver accelerator design:
- the accelerator must deliver an electron beam to the wiggler with the phase space appropriately configured to drive the FEL interaction, and
- it must energy recover the beam after the FEL.

The first requirement reflects the need of the FEL for specific electron beam properties at the wiggler. These, and other relevant system parameters, are presented in Table 1. Particularly important are high charge and a short bunch (to provide high peak current for sufficient FEL gain). The energy recovery requirement alleviates demands on the RF drive system (both installed power and RF window tolerances), limiting cost and reducing radiation power from the dumped drive beam.

Table 1: IR Demo and Upgrade Parameters

| Parameters | Demo | Upgrade | Achieved (5/2000) |
|---|---|---|---|
| energy (MeV) | 35-48 | 80-210 | 20-48 |
| $I_{ave}$ (mA) | 5 | 10 | 5 |
| FEL rep. rate (MHz) | 18.75-75 | 3.9-125 | 18.75-75 |
| $Q_{bunch}$ (pC) | 60 | 135 @ 75 MHz | 135 @ 37.5 MHz |
| bunch $\sigma_l$ (psec) | 0.4 | 0.2 @ 135 pC | 0.4 @ 60 pC |
| $I_{peak}$ (A) | 60 | 270 @ 135 pC | 60 @ 60 pC |
| $\sigma_{\delta p/p}$ (%) | ¼% | ½% | ¼% @ 60 pC |
| $\varepsilon_N$ (mm-mrad) | <13 | <30 | 5-10 @ 60 pC, 25 @ 135 pC |
| $\eta_{FEL}$ | ½% | 1% | >1% |
| $\Delta E_{full}$ after FEL | 5% | 10% | 6-8% |

Both primary requirements imply numerous subsidiary constraints that must be met in the presence of phenomena that can provoke beam quality degradation with consequential poor system performance. These include space charge, BBU, CSR, and the FEL/RF interaction. Amongst the subsidiary constraints are:
- full (6-d) phase space matching at the wiggler,
- longitudinal matching and transverse beam size control during energy recovery,
- preservation of beam quality in the presence of the aforementioned phenomena,
- large acceptance energy recovery transport, and
- insensitivity to, and means for the compensation of, typical accelerator errors such as misalignments, excitation errors and phase errors.

These requirements parallel those imposed on the JLab IR Demo FEL driver; the scope is, however, somewhat more challenging, as inspection of Table 1 implies. Not only must the charge per bunch double, the bunch length at the wiggler must halve so as to quadruple the peak current; moreover, the energy recovery process must accommodate double the momentum spread.

## 2 DESIGN SOLUTION

Figure 1 illustrates a machine design solution meeting the above requirements. It comprises a 10 MeV injector (an upgrade of the existing IR Demo injector from 5 to

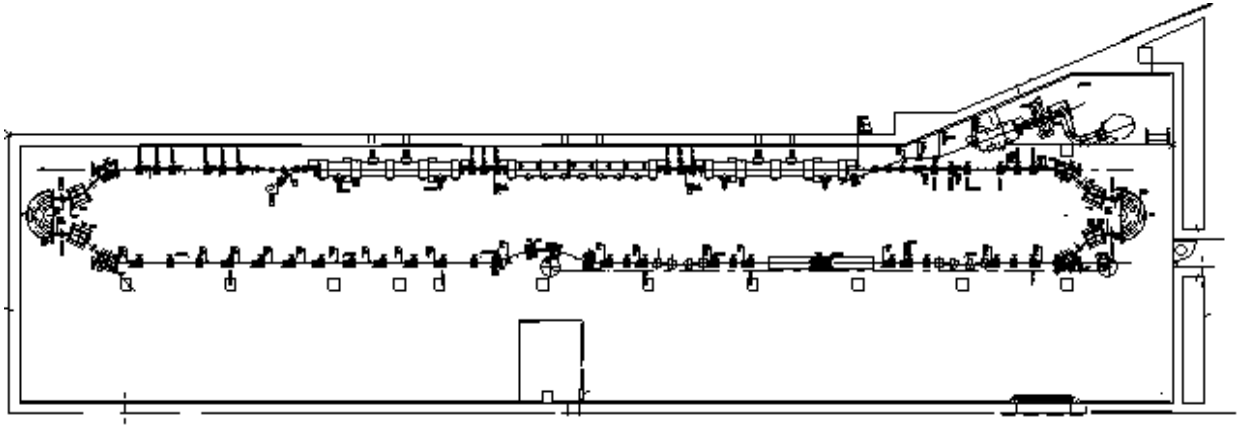

Figure 1: Jefferson Lab 10 kW IR Upgrade FEL. The machine is shown in the facility vault.

10 mA), a linac consisting of three Jefferson Lab cryomodules generating 70 to 200 MeV of energy gain, and a recirculator. The latter provides beam transport to, and phase space conditioning of the accelerated electron beam for, the FEL and then returns and prepares the drive beam for energy recovery in the linac. A summary of the function of each machine section will now be provided.

The injector is a direct upgrade of the IR Demo injector [3] from 5 mA at 10 MeV to 10 mA at 10 MeV. The current will be doubled by an increase of the single bunch charge from 60 pC to 135 pC while maintaining the 75 MHz repetition rate. Adequate source and injector performance has already been demonstrated at the elevated bunch charge at a source-limited repetition rate of 37.5 MHz [4] and will be characterized at full power when a higher power RF drive system is installed.

Transverse and longitudinal phase space management begin in the injector. A four quad telescope matches beam envelopes to the betatron acceptance of the downstream linac, while the choice of RF parameters and momentum compaction of the injection line conspire to produce a long, low momentum spread bunch at the linac.

The linac, which will accelerate the beam by 70 to 200 MeV, comprises three Jefferson Lab cryomodules; the first and third are conventional 5-cell CEBAF designs, the central module is based on new 7-cell Jlab cavities [5]. The lower gradient 5-cell modules are placed at the end of the accelerator to limit RF focussing induced mismatch of the low energy injected and energy recovered beams. Quad triplets between modules provide beam envelope control during both acceleration and energy recovery; the beam is accelerated (energy recovered) off crest (off trough) so as to impose a phase energy correlation on the longitudinal phase space. This is used in subsequent transport to longitudinally match the beam to the required phase space at the wiggler (dump).

The first segment of the recirculator provides transport of the drive beam from the linac to the FEL. Immediately following the linac, a small dipole separates the energy recovered and full energy beams, directing the low energy beam to a dump. A set of three additional identical dipoles completes a chicane of the high energy beam, returning it to the linac axis and directing it through a six quad telescope, which transversely matches it to a recirculation arc. This arc, based on a Bates geometry [6], is used to transport the beam to the machine backleg and to longitudinally condition it in preparation for the FEL interaction. The arc first and second order momentum compactions ($M_{56}$ and $T_{566}$) are selected (using trim quads and sextupoles) to rotate the bunch (which has been slewed by off-crest acceleration) upright at the wiggler and to eliminate phase space curvature, producing a short bunch and high peak current. Downstream of the arc, a FODO transport line conveys the beam to the FEL insertion.

The FEL insertion commences with a chicane around the upstream end of a 32 m $R^5$ high power optical cavity. This provides clearance between electron beam and optical components and generates momentum compaction for a final compression of the bunch length to that required at the wiggler. A six quad telescope transversely matches the beam to the wiggler; a second telescope downstream of the wiggler provides a transverse match from wiggler to the energy recovery transport. A pair of small chicanes embedded in the telescopes allows incorporation of a 16 m broadband optical cavity, which will be used for initial FEL tune-up.

The energy recovery transport consists of a second Bates-style endloop (returning the beam to the linac) followed by a six quad telescope. The beam is matched to the arc by the second telescope of the FEL insertion; the energy recovery telescope matches beam envelopes from the arc to the linac acceptance. Trim quads, sextupoles, and octupoles in the arc adjust momentum compactions through third order to longitudinally rotate the short, very large momentum spread bunch and adjust its curvature and torsion in preparation for energy recovery. As energy recovery occurs off-trough, the imposed phase-energy correlations are selected to generate energy compression during energy recovery, yielding a long, low momentum spread bunch at the dump.

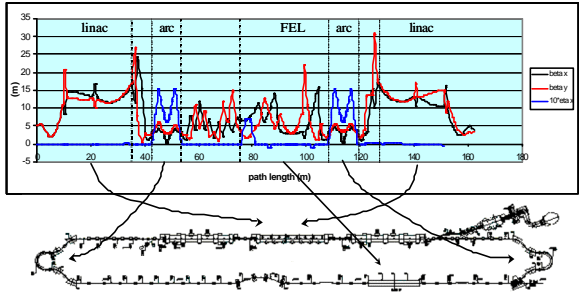

Figure 2: Lattice Functions in IR Upgrade

## 3 SINGLE PARTICLE DYNAMICS

The accelerator design is being evaluated using standard methods. Figure 2 provides the obligatory plot of lattice functions; the chromatic and geometric variations of these and other relevant phase space parameters has been examined for both the linac to wiggler and wiggler to linac transports. The overall system behavior in these initial studies is appropriate for the support of FEL operations. Ray-tracing simulations of energy recovery from wiggler to the linac back-end suggest that IR Demo performance can be duplicated with proper adjustment of the linear, quadratic, and cubic momentum compactions of the Upgrade transport system.

## 4 BEAM DYNAMICS AND COLLECTIVE PHENOMENA

Beam dynamics concerns include configuration of the electron beam for the FEL interaction and the impact of collective effects. Initial studies at 135 pC indicate source performance is adequate. To achieve the higher peak current in the Upgrade, the longitudinal phase space can be managed as in the Demo [7]. Adiabatic damping will reduce the bunch intrinsic relative momentum spread (the "thickness" of the slewed phase ellipse) after the module; when rotated upright at the wiggler, the bunch will then be shorter. Recirculator sextupoles will be used in concert to compensate both lattice and RF-waveform induced curvature of the longitudinal phase space.

A number of collective phenomena can compromise system performance. These include space charge, wake-field effects, BBU, CSR, and the FEL/RF interaction. All are under examination within the context of the machine design presented above. Space charge effects are significant primarily in the injector and at the energy recovery dump; at energies over 25 MeV, IR Demo experience indicates system behavior will be insensitive to space charge driven phenomena. To date, adequate performance at the design single bunch charge of 135 pC has been demonstrated. Further PARMELA and empirical analysis of machine performance, particularly in the injector, will occur as the project continues [8].

A preliminary analysis of wakefield effects [9] suggests an impendence budget based on (2" aperture) IR Demo components will yield marginally acceptable beam performance at the FEL. This is a "worst case" result subject to improvement through use of larger (3") aperture components, which may be allowed following a cost-risk-benefit analysis of the vacuum system design.

Preliminary results extracted from a rudimentary CSR simulation using a 1-dimensional longitudinal wake model suggest Upgrade performance will not be compromised by CSR effects. Future work will involve the application of a more sophisticated CSR model to the design; this will allow comprehensive analysis of CSR effects in the system, including dependence on details of charge distribution within the bunch [10].

Investigation of BBU effects suggests instability thresholds will significantly exceed anticipated operating currents [11]. Study of the FEL/RF interaction suggests that instability in the IR Demo, and thus similarly in the IR Upgrade, is managed by feedback mechanisms in the RF drive system [12].

## 5 PROJECT STATUS

US Navy funding of 9.3 M$ has commenced. A preliminary machine design has been generated and has gone into engineering. Full funding of construction is expected in 2001, with beam operations starting in 2002.

## 6 ACKNOWLEDGEMENTS

This work was supported by the U. S. Dept. of Energy under contract number DE-AC05-84ER40150.